\documentclass{IEEEtran}
\usepackage{cite}
\usepackage{amsmath,amssymb,amsfonts}
\usepackage{algorithmic}
\usepackage{graphicx}
\usepackage{textcomp}
\usepackage{xcolor}
\usepackage{mathtools}
\usepackage{bm}
\def\BibTeX{{\rm B\kern-.05em{\sc i\kern-.025em b}\kern-.08em
    T\kern-.1667em\lower.7ex\hbox{E}\kern-.125emX}}
\begin{document}
\title{\huge{Modeling of Far-Field Quantum Coherence by Dielectric Bodies Based on the Volume Integral Equation Method}}

\author{
	Chengnian Huang, \IEEEmembership{Student Member, IEEE},
	Hangyu Ge,
	Yijia Cheng,
	Zi He, \IEEEmembership{Senior Member, IEEE},
	Feng Liu,
	Wei E. I. Sha, \IEEEmembership{Fellow, IEEE}
	
	\thanks{This work was supported by the National Natural Science Foundation of
		China under Grant 61975177, Grant U20A20164, and Grant U21A6006; in part by the National Key Research and 
		Development Program of China under Grant 2023YFB2806000 and Grant 2022YFA1204700; in part by the Zhejiang Provincial Natural Science Foundation of China under Grant LR25F040001. (Corresponding author:
		Wei E. I. Sha.)}
	\thanks{
		Chengnian Huang, Hangyu Ge, Yijia Cheng, Feng Liu and W. E. I. Sha are with the College of Information Science and Electronic Engineering, Zhejiang University, Hangzhou 310027, China. (e-mail: weisha@zju.edu.cn)}
	\thanks{Zi He is with the Department of Communication Engineering, Nanjing University of Science and Technology, Nanjing 210094, China.}
}



\maketitle

\begin{abstract}	
	The Hong–Ou–Mandel (HOM) effect is a hallmark of nonclassical two-photon interference. This paper develops a unified theory–numerics framework to compute angle-resolved far-field two-photon correlations from arbitrary lossless dielectric scatterers. We describe the input–output relation using a multi-channel scattering formulation that maps two populated incident channels to two selected far-field detection modes, yielding a compact two-channel transfer relation for second-order correlation function and time-domain coincidence counts. The required transfer coefficients are extracted from classical far-field complex amplitudes computed by an fast Fourier transform-accelerated volume integral equation solver, avoiding perfectly matched layers and near-to-far-field post-processing. The method is validated against analytical results for dielectric spheres and demonstrated on a polarization-converting Pancharatnam–Berry-phase metasurface, revealing strong angular dependence of quantum interference and its direct impact on HOM-dip visibility. The framework provides an efficient and physically transparent tool for structure-dependent quantum-correlation analysis, with potential applications in scatterers-enabled quantum state engineering and quantum inverse design.
\end{abstract}


\section{Introduction}
\IEEEPARstart{O}{ver} the past few years, quantum information\cite{blais2020quantuminformation,nielsen2010quantuminformation}, quantum computing\cite{kok2007linearquantumcomputation}, and quantum optics\cite{scully1997quantumoptics} have advanced rapidly, motivating accurate theoretical models and scalable numerical solvers for quantum electromagnetic systems. Within macroscopic quantum electrodynamics (QED)\cite{na2020quantumfdm,chew2024some}, a wide range of effects including spontaneous emission in dielectrics\cite{dung2000spontaneous,scheel1999spontaneous,qiao2011systematic,crenshaw2008comparisondielectric}, Casimir forces\cite{philbin2011casimir}, artificial atoms\cite{bratschitsch2006artificial}, few-photon interference in passive lossless devices\cite{dumke2016roadmap}, and quantum-optical metamaterials\cite{georgi2019metasurface}, can be treated in a unified field-quantization framework.

A central theme in these platforms is scattering-induced quantum interference, which enables controlled manipulation of photon statistics and correlations\cite{slepyan2022modelingAmir}. For single-photon excitation, the detected intensity pattern follows the corresponding classical wave solution for the same input mode. In contrast, under multi-photon excitation, interference between Fock-state probability amplitudes\cite{chew2016quantumfock1,miller2008quantumfock2,gottfried2018quantumfock3} yields genuinely non-classical correlations that do not admit a classical description\cite{lib2022quantumnature}. A canonical example is the Hong-Ou-Mandel (HOM) effect\cite{hong1987measurementhom,prasad1987quantumhom}: two indistinguishable photons impinging on a balanced beam splitter exhibit suppressed coincidences due to destructive two-photon interference.

To predict such non-classical scattering phenomena, both analytical quantization schemes and numerical solvers have been explored. Separation-of-variables quantization has been developed for spherically layered systems and parabolic mirrors\cite{dzsotjan2016modesphere,gutierrez2018photonsmirror}, and far-field correlation functions of canonical dielectric objects can be derived analytically in special geometries\cite{maurer2023quantum}. More generally, Green's-function-based macroscopic QED provides a versatile route for treating structured and inhomogeneous environments\cite{dung1998threegreen,novotny2012principlesgreen,scheel2008macroscopicgreen}.

For arbitrary inhomogeneous and lossless dielectric systems, a common computational route is to expand the quantized field operators on a set of numerical (or quasi-normal) modes, and to promote the classical modal amplitudes to operators by associating them with creation and annihilation operators\cite{scully1997quantumoptics,agarwal2012quantumbeam,bennett2015physicallyoperators,hanson2020aspectsoperators}. For example, characteristic-mode theory\cite{garbacz2005modalmode1,garbacz1971generalizedmode2,garbacz1982antennamode3,harrington1971theorymode4,chen2015characteristicmode6,har1971} has been used to analyze scattering-induced entanglement for perfectly conducting bodies\cite{slepyan2022modelingAmir}. Differential-equation-based solvers \cite{yee1966numericalfdtd,jin2015finitefem,taflove2000computational} have also been employed to compute normal modes and study few-photon interference\cite{na2020quantumfdm,na2020quantumfdtd,savasta2002lightscatter}. However, for large-scale far-field scattering problems, these methods may suffer from numerical dispersion and the need for artificial boundary truncation, which can complicate phase-accurate far-field correlation predictions.

Integral-equation formulations offer an attractive alternative because they leverage analytical Green's functions that satisfy radiation conditions and thus preserve phase propagation exactly. Recent advances by Forestiere \textit{et al.}\cite{forestiere2023integral,forestiere2023static,miano2025quantum} established rigorous integral formulations for macroscopic QED in dispersive dielectric objects. In the lossless setting considered here, radiated observables admit an efficient representation in terms of propagating outgoing channels (e.g., spherical-wave modes), while evanescent components remain confined to the near field and carry no net radiated power. This viewpoint motivates a direct connection between classical far-field scattering amplitudes and quantum photodetection correlations. Building on the framework in\cite{canaguier2016quantumcoherence}, normally ordered photodetection correlation functions can be expressed through classical field quantities, enabling angle-dependent two-photon interference (including HOM-type features) to be evaluated from the far-field responses to independently incident single-photon beams.

In this paper, we study the far-field interference produced when two independent single photons illuminate an arbitrary lossless dielectric object and are correlation-detected at two observation angles. The objective is to compute the second-order normalized correlation function $g^{(2)}$ in both the frequency and time domains by integrating macroscopic quantum theory with a scalable computational electromagnetics (CEM) solver. Specifically, we adopt a volume integral equation (VIE) formulation in which analytical dyadic Green's functions enforce the radiation condition at infinity and avoid numerical dispersion.

For each incident channel, the VIE solver computes the induced polarization currents, from which the far-field response is evaluated directly. This eliminates the need for perfectly matched layers (PMLs) or near-to-far-field post-processing typically required by differential-equation solvers. Moreover, fast Fourier transform (FFT)-accelerated matrix--vector multiplication and block preconditioning enable large-scale simulations of dielectric nanostructures with practical computational cost\cite{huang2023parallel}.

The main contributions are summarized as follows: (1) We develop a unified VIE--dyadic-Green's-function framework that links classical far-field scattering amplitudes to quantum second-order correlation functions for two-photon interference.  (2) The proposed approach avoids heavy normal-mode expansions and mitigates convergence and boundary-related issues commonly encountered in quantum FDTD or mode-based methods, improving efficiency and numerical robustness for large-scale far-field problems.  (3) We connect frequency-domain $g^{(2)}$ maps to time-domain coincidence curves $\tilde N_c(\delta\tau)$, enabling direct prediction of experimentally observable HOM-type interference signatures.

The remainder of this paper is organized as follows. Section~II reviews the quantum-field formalism and derives $g^{(2)}$ in the frequency domain as well as the normalized coincidence function in the time domain. Section~III presents two numerical examples: a dielectric sphere and a polarization-converting metasurface. Section~IV concludes the paper and discusses future directions.\\
\section{Theory}
In the classical world, observable physical quantities such as position or velocity are deterministic. However, due to intrinsic quantum fluctuations, the state of a photon in the microscopic regime must be described by a quantum state vector. For a (non-entangled) single photon prepared as a wavepacket \cite{mandel1995opticalmultimode}, the corresponding state can be expressed as a coherent superposition of single-photon Fock states
\begin{equation}\label{eq1}
	\left|\Psi\right\rangle=\sum_{n=1}^{N} \phi_{n}|1\rangle_{n}
	=\sum_{n=1}^{N} \phi_{n} \hat{a}_{n}^{\dagger}|0\rangle,
	\quad
	\sum_{n=1}^N |\phi_n|^2=1,
\end{equation}
where $\phi_{n}$ denotes the probability amplitude of mode $n$, incorporating the spectral (and, if needed, directional/polarization) profile of the wavepacket; $\hat{a}_{n}^{\dagger}$ is the creation operator, and $|0\rangle$ denotes the vacuum state. The discrete summation over $N$ modes represents a discretization of the wavepacket over an orthonormal set of free-space modes (channels).

If two photons are independent (non-entangled), a convenient two-photon input state can be written as
\begin{equation}\label{eq:twoPhotonState}
	\left|\Psi\right\rangle_{12}
	=
	\left(\sum_{n=1}^{N}\phi_{1n}\hat a_n^\dagger\right)
	\left(\sum_{m=1}^{N}\phi_{2m}\hat a_m^\dagger\right)|0\rangle,
\end{equation}
\begin{equation}
	\sum_n|\phi_{1n}|^2=\sum_m|\phi_{2m}|^2=1,
\end{equation}
where $\phi_{1n}$ and $\phi_{2m}$ describe the two single-photon wavepackets.

Similarly, the classical field quantities in Maxwell's equations are elevated to field operators in the quantum formalism. The electric-field operator is conventionally decomposed into the positive- and negative-frequency components:
\begin{equation}\label{eq2}
	\hat{\mathbf{E}}(\mathbf{r}, t)=\hat{\mathbf{E}}^{(+)}(\mathbf{r}, t)+\hat{\mathbf{E}}^{(-)}(\mathbf{r}, t).
\end{equation}
These components can be expanded over an orthonormal set of free-space modes (channels) as
\begin{equation}\label{eq3}
	\begin{aligned}
		\hat{\mathbf{E}}^{(+)}(\mathbf{r}, t)
		& =\sum_n C\sqrt{\omega_{n}} \mathbf{E}_{n}(\mathbf{r}) \hat{a}_{n}(t) \\
		& =\sum_n C\sqrt{\omega_{n}} \mathbf{E}_{n}(\mathbf{r}) e^{-i \omega_{n} t}\hat{a}_{n},
	\end{aligned}
\end{equation}
\begin{equation}\label{eq4}
	\begin{aligned}
		\hat{\mathbf{E}}^{(-)}(\mathbf{r}, t)
		& =\sum_n C\sqrt{\omega_{n}} \mathbf{E}_{n}^*(\mathbf{r}) \hat{a}_{n}^{\dagger}(t) \\
		& =\sum_n C\sqrt{\omega_{n}} \mathbf{E}_{n}^*(\mathbf{r}) e^{i \omega_{n} t}\hat{a}_{n}^{\dagger},
	\end{aligned}
\end{equation}
where $\omega_{n}$ is the mode frequency, $\mathbf{E}_{n}(\mathbf{r})$ is the corresponding (normalized) mode profile, and $C$ is a normalization constant fixed by the chosen quantization convention. $\hat{a}_{n}$ and $\hat{a}_{n}^{\dagger}$ are the annihilation and creation operators, respectively.

\begin{figure}[!t]
	\centering
	\includegraphics[height=0.24\textwidth,width=0.75\columnwidth]{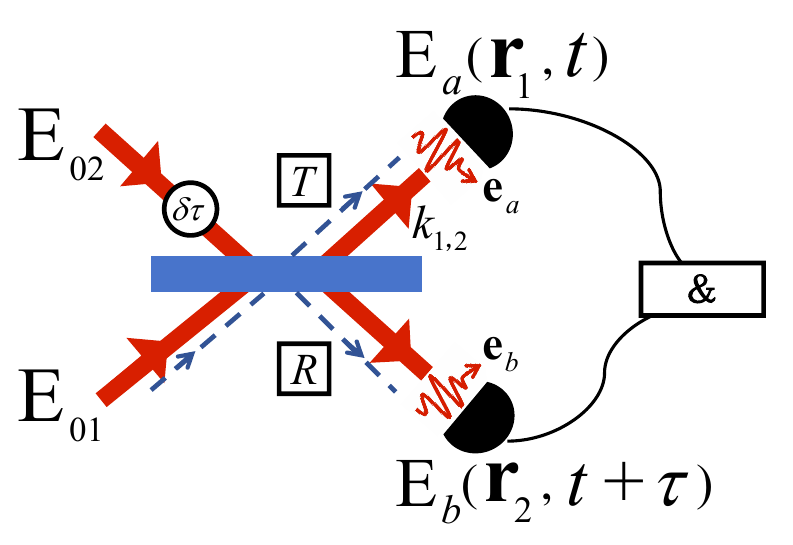}
	\caption{The schematic of the photon detection setup, where two detectors are placed at positions $\mathbf{r}_1$ and $\mathbf{r}_2$, with polarization detection directions $\mathbf{e}_a$ and $\mathbf{e}_b$, respectively.}
	\label{fig1}
\end{figure}

In order to explore the quantum characteristics of electromagnetic fields, we consider the probability of joint photodetection, which is mathematically modeled by correlation functions. The photodetection scheme with two detectors located at $\mathbf{r}_1$ and $\mathbf{r}_2$ is illustrated in Fig.~\ref{fig1}.

The first-order correlation function, proportional to the single-photon detection probability at position $\mathbf{r}_1$ with polarization $\mathbf{e}_a$, is defined as
\begin{equation}\label{eq5}
	G^{(1)}_{a}\left(\mathbf{r}_{1}, t\right)=
	\langle\Psi|\hat{E}^{(-)}_{a}\left(\mathbf{r}_1,t\right)\hat{E}^{(+)}_{a}\left(\mathbf{r}_1,t\right)|\Psi\rangle,
\end{equation}
where $\hat{E}^{(-)}_{a}\left(\mathbf{r}_1,t\right)=\hat{\mathbf{E}}^{(-)}\left(\mathbf{r}_1,t\right)\cdot\mathbf{e}_a=
\left[\hat{E}^{(+)}_{a}\left(\mathbf{r}_1,t\right)\right]^{\dagger}$ is the projection of the negative-frequency field operator onto the polarization vector.

The second-order correlation function, corresponding to fourth-order field interference, yields the joint probability to detect one photon at $\mathbf{r}_1$ with polarization $\mathbf{e}_a$ and another photon at $\mathbf{r}_2$ with polarization $\mathbf{e}_b$:
\begin{equation}\label{eq6}
	\begin{aligned}
		G^{(2)}_{ab}(\mathbf{r}_1, \mathbf{r}_2, t)
		= \langle \Psi | &
		\hat{E}^{-}_{a}(\mathbf{r}_1, t)
		\hat{E}^{-}_{b}(\mathbf{r}_2, t)
		\\[4pt]
		&\times
		\hat{E}^{+}_{b}(\mathbf{r}_2, t)
		\hat{E}^{+}_{a}(\mathbf{r}_1, t)
		| \Psi \rangle.
	\end{aligned}
\end{equation}
To assess the degree of quantum coherence, the normalized second-order correlation function \cite{brown1957interferometryg2,glauber1963quantumg2} is introduced:
\begin{equation}\label{eq7}
	\begin{aligned}
		g^{(2)}_{ab}\left(\mathbf{r}_{1}, \mathbf{r}_{2},t\right)
		=
		\frac{G^{(2)}_{ab}\left(\mathbf{r}_{1},\mathbf{r}_{2}, t\right)}
		{G^{(1)}_{a}\left(\mathbf{r}_{1}, t\right)G^{(1)}_{b}\left(\mathbf{r}_{2}, t\right)}.
	\end{aligned}
\end{equation}
\subsection{The correlation function in the frequency domain}
For a linear, time-invariant, and lossless dielectric scatterer, the input-output relation of the quantized field can be formulated as a unitary mapping between complete orthonormal sets of asymptotic modes (channels). We start from a plane-wave basis (including polarization) for the incident field, with annihilation operators $\{\hat{a}_{\alpha}\}$ satisfying the canonical commutation relations $[\hat{a}_{\alpha},\hat{a}^{\dagger}_{\beta}]=\delta_{\alpha\beta}$. For far-field analysis, it is convenient to express the same free-space field in an incoming spherical-wave basis $\{\hat{b}^{\mathrm{(in)}}_{\mu}\}$. These two complete bases are related by a frequency-dependent unitary transformation,
\begin{equation}\label{eq:pw2sw}
	\hat{b}^{\mathrm{(in)}}_{\mu}=\sum_{\alpha} U_{\mu\alpha}\,\hat{a}_{\alpha},\quad U^{\dagger}U=I,
\end{equation}
which preserves commutation relations.

The scatterer maps incoming channels into outgoing channels through the scattering matrix
\begin{equation}\label{eq:SmatrixFull}
	\hat{b}^{\mathrm{(out)}}_{\nu}=\sum_{\mu} S_{\nu\mu}\,\hat{b}^{\mathrm{(in)}}_{\mu}.
\end{equation}
For a lossless object, the full multi-channel matrix $S$ is unitary ($S^{\dagger}S=I$), ensuring preservation of the bosonic commutators for the output operators $\{\hat{b}^{\mathrm{(out)}}_{\nu}\}$.

In this work, only two independent single-photon wavepackets are injected into two selected input channels (denoted as $1$ and $2$), while all other input channels are in the vacuum state. Because the photodetection correlations considered here are normally ordered, vacuum input channels contribute zero to the relevant expectation values and can therefore be omitted when evaluating $G^{(1)}$ and $G^{(2)}$.

Under a narrowband approximation (the system response varies negligibly over the photon bandwidth around $\omega_0$)\footnote{Regarding the source characteristics, the incident single-photon wavepacket generated by high-performance quantum emitters is assumed, such as the wavelength-tunable semiconductor quantum dots recently demonstrated in \cite{chen2024wavelength}.}, the detected positive-frequency field operators at the two detectors can be written as linear combinations of the two excited input operators, with the corresponding transfer coefficients treated as approximately frequency-independent over that bandwidth:
\begin{equation}\label{eq:field_transfer1}
	\hat{E}^{(+)}_{a}(\mathbf{r}_1,t)=\sum_{j=1}^{2} T_{aj}\,e^{-i\omega_0 t}\,\hat a_j,
\end{equation}
\begin{equation}\label{eq:field_transfer2}
	\hat{E}^{(+)}_{b}(\mathbf{r}_2,t)=\sum_{j=1}^{2} T_{bj}\,e^{-i\omega_0 t}\,\hat a_j,
\end{equation}
where $T_{aj}$ and $T_{bj}$ are complex transfer coefficients that incorporate the scatterer response, propagation, and polarization projection. These coefficients are extracted from the full-wave (VIE-FFT) solution of the corresponding classical scattering problem.
Note that the $2\times2$ matrix formed by $\{T_{aj},T_{bj}\}$ is a projection of the full multi-channel mapping onto two selected detection modes and is therefore not generally unitary, because power may scatter into unobserved output channels; unitarity holds only for the full multi-channel $S$-matrix.

Equivalently, keeping explicit mode functions as in (\ref{eq3})--(\ref{eq4}), one may write
\begin{equation}\label{eq8}
	\begin{aligned}
		\hat{E}^{+}_{a}(\mathbf{r}_1,t) 
		&= C \sqrt{\omega_{k_{1}}}\, E_{ak_{1}}(\mathbf{r}_1)\, e^{-i \omega_{k_{1}} t}\, \hat{a}_{1} \\
		&\quad + C \sqrt{\omega_{k_{2}}}\, E_{ak_{2}}(\mathbf{r}_1)\, e^{-i \omega_{k_{2}} t}\, \hat{a}_{2}
	\end{aligned}
\end{equation}
\begin{equation}\label{eq9}
	\begin{aligned}
		\hat{E}^{+}_{b}(\mathbf{r}_2,t) 
		&= C \sqrt{\omega_{k_{1}}}\, E_{bk_{1}}(\mathbf{r}_2)\, e^{-i \omega_{k_{1}} t}\, \hat{a}_{1} \\
		&\quad + C \sqrt{\omega_{k_{2}}}\, E_{bk_{2}}(\mathbf{r}_2)\, e^{-i \omega_{k_{2}} t}\, \hat{a}_{2}.
	\end{aligned}
\end{equation}
The negative-frequency components follow by Hermitian conjugation.

To incorporate a finite numerical aperture (NA), we define an aperture-collected field operator
\begin{equation}\label{eq:NAfield}
	\hat E^{(+)}_{a,\mathrm{NA}}(t)=\int_{\mathrm{NA}} W_a(\mathbf{k})\,\hat E^{(+)}(\mathbf{k},t)\,d^2\mathbf{k},
\end{equation}
and similarly for $\hat E^{(+)}_{b,\mathrm{NA}}(t)$, where $W(\mathbf{k})$ is the angular sensitivity function of the detector. The experimentally relevant quantities are then
$G^{(1)}_{a,\mathrm{NA}}=\langle \hat E^{(-)}_{a,\mathrm{NA}}\hat E^{(+)}_{a,\mathrm{NA}}\rangle$,
$G^{(2)}_{ab,\mathrm{NA}}=\langle \hat E^{(-)}_{a,\mathrm{NA}}\hat E^{(-)}_{b,\mathrm{NA}}\hat E^{(+)}_{b,\mathrm{NA}}\hat E^{(+)}_{a,\mathrm{NA}}\rangle$,
followed by $g^{(2)}_{ab,\mathrm{NA}}=G^{(2)}_{ab,\mathrm{NA}}/(G^{(1)}_{a,\mathrm{NA}}G^{(1)}_{b,\mathrm{NA}})$. Note that in our numerical implementation, under the narrowband approximation, the spatial integration over the numerical aperture is not explicitly performed. Instead, by approximating the collected signal through contributions from specific discrete modes at the center of the detection angles, the substitution of (\ref{eq:field_transfer1}--\ref{eq:field_transfer2}) into (\ref{eq7}) yeilds:
\begin{equation}\label{eq11}
	\begin{aligned}
		g^{(2)}_{ab}=\frac{\smashoperator{\sum\limits_{m, n, p, q\in\{1,2\}}} T_{a m}^{*} T_{b n}^{*} T_{b p} T_{a q}\left\langle \Psi \right| \hat{a}_{m}^{\dagger} \hat{a}_{n}^{\dagger} \hat{a}_{p} \hat{a}_{q}\left|\Psi \right\rangle}{\smashoperator{\sum\limits_{m, q \in\{1,2\}}}T_{a m}^{*} T_{a q} \langle\Psi| \hat{a}_{m}^{\dagger} \hat{a}_{q}|\Psi\rangle\smashoperator{\sum\limits_{n, p \in\{1,2\}}} T_{b n}^{*} T_{b p} \langle\Psi| \hat{a}_{n}^{\dagger} \hat{a}_{p}|\Psi\rangle}\\
	\end{aligned}.
\end{equation}

For the two-single-photon Fock input $|1_1,1_2\rangle=\hat a_1^\dagger\hat a_2^\dagger|0\rangle$, the fourth-order expectation
$\langle \hat a_m^\dagger \hat a_n^\dagger \hat a_p \hat a_q\rangle$ vanishes whenever any annihilation operator attempts to remove two photons from the same occupied mode (since $n_1=n_2=1$). Therefore, only terms that annihilate one photon from each occupied mode survive, resulting in the direct and exchange two-photon paths and their interference.
If $\omega_{k_1}\neq\omega_{k_2}$, the cross terms acquire a beating phase $e^{-i(\omega_{k_1}-\omega_{k_2})t}$ and vanish under experimental time averaging unless the two wavepackets are sufficiently overlapped in frequency (i.e., effectively indistinguishable). In the degenerate (or effectively narrowband) case $\omega_{k_1}\approx\omega_{k_2}$, this explicit time dependence cancels in $g^{(2)}$.

In the degenerate/narrowband case and using (\ref{eq8})--(\ref{eq9}), the normalized correlation reduces to a compact far-field form:
\begin{equation}\label{eq10}
	\begin{aligned}
		&g^{(2)}_{ab}(\mathbf{r}_{1}, \mathbf{r}_{2})\\ 
		&= \frac{
			\omega_{k_1} \omega_{k_2} \left| 
			E_{a k_1}(\mathbf{r}_1) E_{b k_2}(\mathbf{r}_2) 
			+ E_{a k_2}(\mathbf{r}_1) E_{b k_1}(\mathbf{r}_2) 
			\right|^2
		}{
			\left( \sum_{l=1,2} \omega_{k_l} \left| E_{a k_l}(\mathbf{r}_1) \right|^2 \right)
			\left( \sum_{l=1,2} \omega_{k_l} \left| E_{b k_l}(\mathbf{r}_2) \right|^2 \right)
		}
	\end{aligned}
\end{equation}
The numerator contains the classical-like contributions from the direct and exchange paths and, crucially, the cross terms that arise from quantum indistinguishability. When the two photons are perfectly indistinguishable, these cross terms yield maximum two-photon interference (e.g., the Hong-Ou-Mandel suppression of coincidences); when they are fully distinguishable, the cross terms vanish and the coincidence reduces to an incoherent sum of independent intensities. Ultimately, for the two-single-photon Fock input considered here and the above normalization, the resulting $g^{(2)}$ is bounded between 0 and 1, reflecting the antibunching associated with single-photon excitation \cite{kimble1977photonbunching,messin2001bunching}. In contrast, incoherent classical emitters can exhibit autocorrelation values exceeding unity under the conditions $\textbf{r}_1=\textbf{r}_2$ and $\textbf{e}_a=\textbf{e}_b$\cite{carminati2015speckle}.

\subsection{The normalized coincidence number function in the time domain}
To model experimentally observed coincidence measurements in the time domain, we derive the expected coincidence number $N_c$ as a function of an externally controlled optical path delay $\delta\tau$, following the standard treatment in \cite{hong1987measurementhom}. The delay $\delta\tau$ represents a controllable input delay, effectively shifting the arrival time of one wavepacket relative to the other.

\subsubsection{Beam-splitter reference formulation}
We first recall the phase-asymmetric beam-splitter model. Let $\hat E_{01}^{(+)}(t)$ and $\hat E_{02}^{(+)}(t)$ denote the incident positive-frequency field operators associated with the two input channels. An adjustable delay $\delta\tau$ is applied to the second input wavepacket. The output fields at the two detectors are then written as linear combinations of the delayed inputs:
\begin{equation}\label{eq15}
	\hat{E}_a^{(+)}(\mathbf{r}_1,t)=\sqrt{T}\,\hat{E}_{01}^{(+)}(t)+i\sqrt{R}\,\hat{E}_{02}^{(+)}(t+\delta\tau),
\end{equation}
\begin{equation}\label{eq16}
	\hat{E}_b^{(+)}(\mathbf{r}_2,t+\tau)=\sqrt{T}\,\hat{E}_{02}^{(+)}(t+\tau)+i\sqrt{R}\,\hat{E}_{01}^{(+)}(t+\tau-\delta\tau),
\end{equation}
where $\tau\equiv t_b-t_a$ is the detection-time difference between the two detectors. Note that $\delta\tau$ is a controlled input delay applied to one wavepacket, whereas $\tau$ is the random detection time difference over which coincidence events are accumulated.

The second-order correlation function $G^{(2)}_{ab}(\tau;\delta\tau)$ is obtained from (\ref{eq6}) by substituting the field operators above. For two independent single-photon wavepackets with spectral amplitude functions $\phi_1(\omega)$ and $\phi_2(\omega)$, the key quantity governing two-photon interference is the wavepacket overlap function
\begin{equation}\label{eq:h_def}
	h(\tau)\;=\;\int_{-\infty}^{\infty} d\omega\; \phi_1^*(\omega)\,\phi_2(\omega)\,e^{-i\omega\tau}.
\end{equation}
For identical Gaussian spectra centered at $\omega_0$ with rms bandwidth $1/\sigma$, one obtains (up to an irrelevant global phase)
\begin{equation}\label{eq:h_gauss}
	h(\tau)=\exp\!\left(-\frac{\tau^2}{2\sigma^2}\right)\,e^{-i\omega_0\tau},
\end{equation}
and consequently $|h(\tau)|^2=\exp(-\tau^2/\sigma^2)$.

With this definition, the beam-splitter correlation function takes the standard HOM form
\begin{equation}\label{eq17}
	\begin{aligned}
		G^{(2)}_{ab}(\tau;\delta\tau)
		&=K\Big[
		T^{2}|h(\tau)|^{2}
		+R^{2}|h(2\delta\tau-\tau)|^{2} \\
		&-RT\,h^{*}(\tau)\,h(2\delta\tau-\tau)
		-RT\,h(\tau)\,h^{*}(2\delta\tau-\tau)
		\Big],
	\end{aligned}
\end{equation}
where $K$ is a positive constant determined by the overall normalization and detector response. In typical coincidence experiments, the integration window is much longer than the coherence time, so that the expected coincidence number is
\begin{equation}\label{eq:Nc_def}
	N_c(\delta\tau)=\int_{-\infty}^{\infty} G^{(2)}_{ab}(\tau;\delta\tau)\,d\tau.
\end{equation}

\subsubsection{Extension to an arbitrary lossless scattering environment}
We now generalize the above beam-splitter model to an arbitrary (lossless) scattering environment computed by the full-wave solver. The environment determines how the two incident wavepackets are mapped into far-field observation directions and polarizations at $\mathbf r_1$ and $\mathbf r_2$. In the narrowband regime, the detector-projected far-field amplitudes at $\omega_0$ play the role of complex transfer coefficients. Specifically, the two indistinguishable two-photon detection paths are associated with the complex amplitudes
\begin{equation}\label{eq:path_amplitudes}
	A_{\mathrm{dir}} = E_{a k_1}(\mathbf r_1)\,E_{b k_2}(\mathbf r_2),
	\quad
	A_{\mathrm{ex}}  = E_{a k_2}(\mathbf r_1)\,E_{b k_1}(\mathbf r_2),
\end{equation}
evaluated at $\omega_{k_1}=\omega_{k_2}=\omega_0$, while the temporal/spectral profile of the wavepackets is retained through $h(\cdot)$. This operator-level mapping from input channels to far-field detection modes follows from the Heisenberg evolution in linear media and can be formulated via Green's-function quantization in inhomogeneous environments \cite{chew1999waves,knoll2000qed,viviescas2003field}.

Under this framework, the time-resolved correlation function can be written as
\begin{equation}\label{eq:G2_scatter}
	\begin{aligned}
		&G^{(2)}_{ab}(\tau;\delta\tau)	=
		K\Big[
		|A_{\mathrm{dir}}|^2\,|h(\tau)|^2
		+|A_{\mathrm{ex}}|^2\,|h(2\delta\tau-\tau)|^2 \\
		&+A_{\mathrm{dir}}A_{\mathrm{ex}}^{*}\,h(\tau)\,h^{*}(2\delta\tau-\tau)
		+A_{\mathrm{dir}}^{*}A_{\mathrm{ex}}\,h^{*}(\tau)\,h(2\delta\tau-\tau)
		\Big].
	\end{aligned}
\end{equation}
Equation (\ref{eq:G2_scatter}) reduces to (\ref{eq17}) for an ideal beam splitter when $A_{\mathrm{dir}}=T$ and $A_{\mathrm{ex}}=-R$.

\subsubsection{Normalized coincidence number}
Experimentally, one is typically interested in the coincidence number integrated over $\tau$. Using (\ref{eq:Nc_def}) and (\ref{eq:G2_scatter}), we define the normalized coincidence number $\tilde N_c(\delta\tau)$ such that $\tilde N_c(\delta\tau)\to 1$ as $|\delta\tau|\to\infty$:
\begin{equation}\label{eq18}
	\tilde N_c(\delta\tau)
	=
	\frac{N_c(\delta\tau)}{D_c},
\end{equation}
where the normalization factor $D_c$ is chosen as the integral of the product of two terms (i.e., the large-delay limit of the coincidence rate),
\begin{equation}\label{eq:Dc_def}
	D_c
	=
	\int_{-\infty}^{\infty} G^{(1)}_{a}(\tau;\delta\tau)\,
	G^{(1)}_{b}(\tau;\delta\tau)\,d\tau
\end{equation}
so that $\tilde N_c(\delta\tau)$ is dimensionless and asymptotically approaches unity.

Carrying out the $\tau$-integration in (\ref{eq:G2_scatter}) yields a compact decomposition
\begin{equation}\label{eq:Nc_expand}
	\tilde N_c(\delta\tau)
	=
	\frac{N_1(\delta\tau)+N_2(\delta\tau)+N_3(\delta\tau)+N_4(\delta\tau)}{D_c},
\end{equation}
with
\begin{equation}\label{eq19}
	N_1(\delta\tau)
	=
	|A_{\mathrm{dir}}|^{2}
	\int_{-\infty}^{\infty} |h(\tau)|^{2}\, d\tau,
\end{equation}
\begin{equation}\label{eq20}
	N_2(\delta\tau)
	=
	|A_{\mathrm{ex}}|^{2}
	\int_{-\infty}^{\infty} \left| h(2\delta\tau-\tau) \right|^{2}\, d\tau,
\end{equation}
\begin{equation}\label{eq21}
	N_3(\delta\tau)
	=
	A_{\mathrm{dir}}A_{\mathrm{ex}}^{*}
	\int_{-\infty}^{\infty} h(\tau)\, h^{*}(2\delta\tau-\tau)\, d\tau,
\end{equation}
\begin{equation}\label{eq22}
	N_4(\delta\tau)
	=
	A_{\mathrm{dir}}^{*}A_{\mathrm{ex}}
	\int_{-\infty}^{\infty} h^{*}(\tau)\, h(2\delta\tau-\tau)\, d\tau.
\end{equation}
Moreover, for identical Gaussian spectra (\ref{eq:h_gauss}), one obtains the standard HOM envelope:
\begin{equation}\label{eq:HOM_envelope}
	\int_{-\infty}^{\infty} h(\tau)\, h^{*}(2\delta\tau-\tau)\, d\tau
	\;\propto\;
	\exp\!\left(-\frac{\delta\tau^2}{\sigma^2}\right),
\end{equation}
so that the dip width is governed by the coherence time, represented by the Gaussian width parameter $\sigma$. It should be explicitly clarified that $\sigma$ is an assumed constant, independent of the monochromatic frequency-domain scattering results. This parameter is employed solely to enable the transition from monochromatic data to a temporal profile; thus, the resulting coincidence curves do not represent a full broadband frequency-domain simulation, but rather a theoretical visualization based on the assumed temporal envelope.

The temporal envelope of the excitation pulses thus defines the coherence time of the generated photons and determines the characteristic width of the interference curve. In this work, we emphasize the comparative behavior of $\tilde N_c(\delta\tau)$ across different scattering environments; the absolute temporal scale can be mapped to a specific experimental implementation through the pulse bandwidth and detector timing response. Accordingly, $\sigma$ is assigned as a dimensionless value consistent with this normalized temporal scale.
	 
\subsection{Numerical method}
To obtain the far-field complex amplitudes at the detection directions for a given frequency, we employ a VIE solver accelerated by FFT in this work. Following the electromagnetic equivalence principle, the dielectric scatterer occupying a volume $V$ is embedded in a homogeneous background (vacuum), and the scattering effect is represented by an induced volumetric polarization current inside $V$. In the frequency domain ($e^{-i\omega t}$ convention), the polarization current is defined as
\begin{equation}\label{eq:Js_def}
	\mathbf{J}^{\mathrm{s}}(\mathbf r)= -i\omega\left[\varepsilon(\mathbf r)-\varepsilon_0\right]\mathbf E^{\mathrm{tot}}(\mathbf r),
	\quad \mathbf r\in V,
\end{equation}
where $\varepsilon(\mathbf r)$ is the (possibly piecewise-constant) permittivity distribution and $\mathbf E^{\mathrm{tot}}$ is the total electric field. The scattered field produced by $\mathbf J^{\mathrm s}$ is expressed via the free-space dyadic Green's function $\overleftrightarrow{\mathbf G}$ as
\begin{equation}\label{eq:Esca_from_J}
	\mathbf E^{\mathrm{sca}}(\mathbf r)=i\omega\mu_0\int_V \overleftrightarrow{\mathbf G}(\mathbf r,\mathbf r')\cdot \mathbf J^{\mathrm s}(\mathbf r')\,dv',
\end{equation}
and the total field satisfies $\mathbf E^{\mathrm{tot}}=\mathbf E^{\mathrm{inc}}+\mathbf E^{\mathrm{sca}}$. Substituting (\ref{eq:Js_def})--(\ref{eq:Esca_from_J}) yields a VIE in terms of the unknown current $\mathbf J^{\mathrm s}$:
\begin{equation}\label{eq24}
	\begin{aligned}
	-\frac{\mathbf{J}^{\mathrm{s}}(\mathbf{r})}{i \omega\left(\varepsilon(\mathbf r)-\varepsilon_0\right)}
	-i \omega \mu_0 \int_V \overleftrightarrow{\mathbf{G}}\left(\mathbf{r}, \mathbf{r}^{\prime}\right)\cdot \mathbf{J}^{\mathrm{s}}\left(\mathbf{r}^{\prime}\right) d v^{\prime}
	=\mathbf{E}^{\mathrm{inc}}(\mathbf r),\\
	 \mathbf r\in V,
	\end{aligned}
\end{equation}
where the free-space dyadic Green’s function is
\begin{equation}\label{eq:dyadicG}
	\overleftrightarrow{\mathbf{G}}\left(\mathbf{r}, \mathbf{r}^{\prime}\right)
	=\left(\overleftrightarrow{\mathbf{I}}+\frac{1}{k_{0}^2}\nabla\nabla\right)
	\frac{e^{i k_0\left|\mathbf{r}-\mathbf{r}^{\prime}\right|}}{4 \pi\left|\mathbf{r}-\mathbf{r}^{\prime}\right|},
\end{equation}
with $k_0=\omega\sqrt{\mu_0\varepsilon_0}$, and $\varepsilon_0$ and $\mu_0$ the permittivity and permeability of free space.

We discretize (\ref{eq24}) using the Method of Moments (MoM) \cite{harrington1993field}. The unknown volumetric current is expanded on a uniform hexahedral grid using rooftop (piecewise-linear) basis functions oriented along the three Cartesian directions:
\begin{equation}\label{eq:Js_expand}
	\mathbf{J}^{\mathrm{s}}(\mathbf r)
	=
	\sum_{i\in\{x,y,z\}}\hat{\mathbf p}_i
	\sum_{k=1}^{N_{1}}\sum_{m=1}^{N_{2}}\sum_{n=1}^{N_{3}}
	J_i(k,m,n)\,T^{(i)}_{k m n}(\mathbf r),
\end{equation}
where $\hat{\mathbf p}_i$ denotes the unit vector along direction $i$, $J_i(k,m,n)$ are the expansion coefficients, and $T^{(i)}_{k m n}$ are the volumetric rooftop basis functions. Pulse (piecewise-constant) testing functions are used in a Galerkin-like collocation manner. The rooftop expansion is advantageous because it enforces appropriate inter-cell continuity (notably the normal flux continuity across shared faces), improving numerical stability and accuracy for volumetric formulations.

For an arbitrary geometry, a structured (uniform) voxelization leads to a discrete convolution structure between the current unknowns and the Green’s function kernel. Consequently, matrix-vector multiplications (MVMs) can be accelerated by FFT, reducing the computational cost from $\mathcal O(N^2)$ to approximately $\mathcal O(N\log N)$ for large-scale problems. The resulting linear system can be efficiently solved using the bi-conjugate gradient stabilized (BiCGStab) iterative method \cite{van1992bi}. For large-scale scatterers, additional efficiency gains can be achieved via block preconditioners and parallel computing \cite{huang2023parallel}.

Once the polarization current is obtained, the far-field scattered electric field can be evaluated using the radiation-zone asymptotic form of the dyadic Green’s function. The scattered far-field amplitude in the observation direction $\hat{\mathbf r}$ can be written as
\begin{equation}\label{eq25}
	\mathbf{E}^{\mathrm{sca}}_{\infty}(\hat{\mathbf{r}})
	=
	C_{\infty}\int_V
	\left[\hat{\mathbf{r}}\times\left(\mathbf{J}^{\mathrm{s}}(\mathbf{r}^{\prime})\times \hat{\mathbf{r}}\right)\right]
	e^{-i k_0 \hat{\mathbf{r}} \cdot \mathbf{r}^{\prime}}
	\, d \mathbf{r}^{\prime},
\end{equation}
where $C_{\infty}$ is a frequency-dependent proportionality constant determined by the chosen far-field normalization, and $\hat{\mathbf r}$ denotes the unit vector in the observation direction. The detector-projected complex amplitudes used in the correlation analysis are then obtained as
\begin{equation}\label{eq:Eproj}
	E_{a k}(\mathbf r_1)= \mathbf e_a\cdot \mathbf E^{\mathrm{tot}}_{\infty}(\hat{\mathbf r}_1;\omega_k),\enspace
	E_{b k}(\mathbf r_2)= \mathbf e_b\cdot \mathbf E^{\mathrm{tot}}_{\infty}(\hat{\mathbf r}_2;\omega_k),
\end{equation}
where $\mathbf E^{\mathrm{tot}}_{\infty}=\mathbf E^{\mathrm{inc}}_{\infty}+\mathbf E^{\mathrm{sca}}_{\infty}$ and $\hat{\mathbf r}_{1,2}$ correspond to the two detection angles.

Finally, the geometric features of different structures lead to distinct spatial distributions of $\mathbf{J}^{\mathrm{s}}$, which directly dictate the angular dependence of the far-field amplitudes (\ref{eq25}) and thereby control the interference behavior in the two-photon correlation functions reported in this work.

\section{Results}
Following the theoretical derivations, the normalized two-photon correlation observables are fully determined by the complex far-field amplitudes at the detection directions. In the regime considered here, each incident single-photon wavepacket occupies a well-defined input spatial mode (channel) with a narrowband spectral envelope. For normally ordered photodetection correlations, the dependence on the quantum state enters only through the occupied input-mode operators, while the scattering from input channels to far-field detection modes is strictly linear. Therefore, the required transfer coefficients can be extracted from classical frequency-domain scattering simulations: for each excited input channel (incident direction and polarization), we compute the corresponding complex far-field at the detector directions, and then evaluate $g^{(2)}$ and $\tilde N_c$ using the closed-form expressions in Sec.~II.

In this section, two representative scenarios are investigated: (i) two-photon scattering from a dielectric sphere, which exhibits angle-dependent HOM-type suppression of coincidences; and (ii) the generation of spatial--polarization entanglement mediated by a space-variant Pancharatnam--Berry phase metasurface. In both cases, the structure acts as a distributed three-dimensional beam splitter: the effective ``path amplitudes'' are encoded in the continuous angular dependence of the far-field response rather than in discrete optical paths. Specifically, $\sigma = 1.2$ is adopted to facilitate the temporal visualization of the HOM dip. This choice ensures a clear representation of the coincidence features within the chosen time window, independent of the frequency-domain scattering calculations.

\subsection{Dielectric Sphere}
To study the joint detection probability, we consider two independent single-photon wavepackets injected into two orthogonal input channels: one is a $y$-polarized plane wave propagating along the $x$-direction ($\mathbf{k}_1=k\mathbf{e}_x$), and the other is an $x$-polarized plane wave propagating along the $y$-direction ($\mathbf{k}_2=k\mathbf{e}_y$). The detectors are placed in the far field at directions $(\theta_1,\phi_1)$ and $(\theta_2,\phi_2)$, respectively.

To suppress the dominant co-polarized incident contribution in the detected signal and to isolate the scattered-field-induced correlations, we detect only the $z$-polarized component (i.e., $a=b=z$), following \cite{maurer2023quantum}. In the narrowband (degenerate) regime $\omega_{\mathbf{k}_1}\approx\omega_{\mathbf{k}_2}=\omega_0$, \eqref{eq10} reduces to
\begin{equation}\label{eq26}
	\begin{aligned}
		&g_{zz}^{(2)}\left(\mathbf{r}_1, \mathbf{r}_2\right)\\
		&=\frac{\omega_{\textit{k}_1} \omega_{\textit{k}_2}\left|E_{z \textit{k}_1}\left(\mathbf{r}_1\right) E_{z \textit{k}_2}\left(\mathbf{r}_2\right)+E_{z \textit{k}_2}\left(\mathbf{r}_1\right) E_{z \textit{k}_1}\left(\mathbf{r}_2\right)\right|^2}{\sum_{l=1,2} \omega_{\textit{k}_l}\left|E_{z \textit{k}_l}\left(\mathbf{r}_1\right)\right|^2 \sum_{l=1,2} \omega_{\textit{k}_l}\left|E_{z \textit{k}_l}\left(\mathbf{r}_2\right)\right|^2}. 
	\end{aligned}
\end{equation}
Here $E_{z\mathbf{k}_j}(\mathbf r)$ denotes the detector-projected far-field complex amplitude at the observation direction corresponding to detector location $\mathbf r$, generated by the $j$-th incident channel. Equation (\ref{eq26}) explicitly shows that $g_{zz}^{(2)}$ is governed by the interference between the direct and exchange two-photon amplitudes, which vary continuously with observation angles.

\subsubsection{Validation against analytical reference}
\begin{figure}[!t]
	\centering
	\includegraphics[width=\columnwidth]{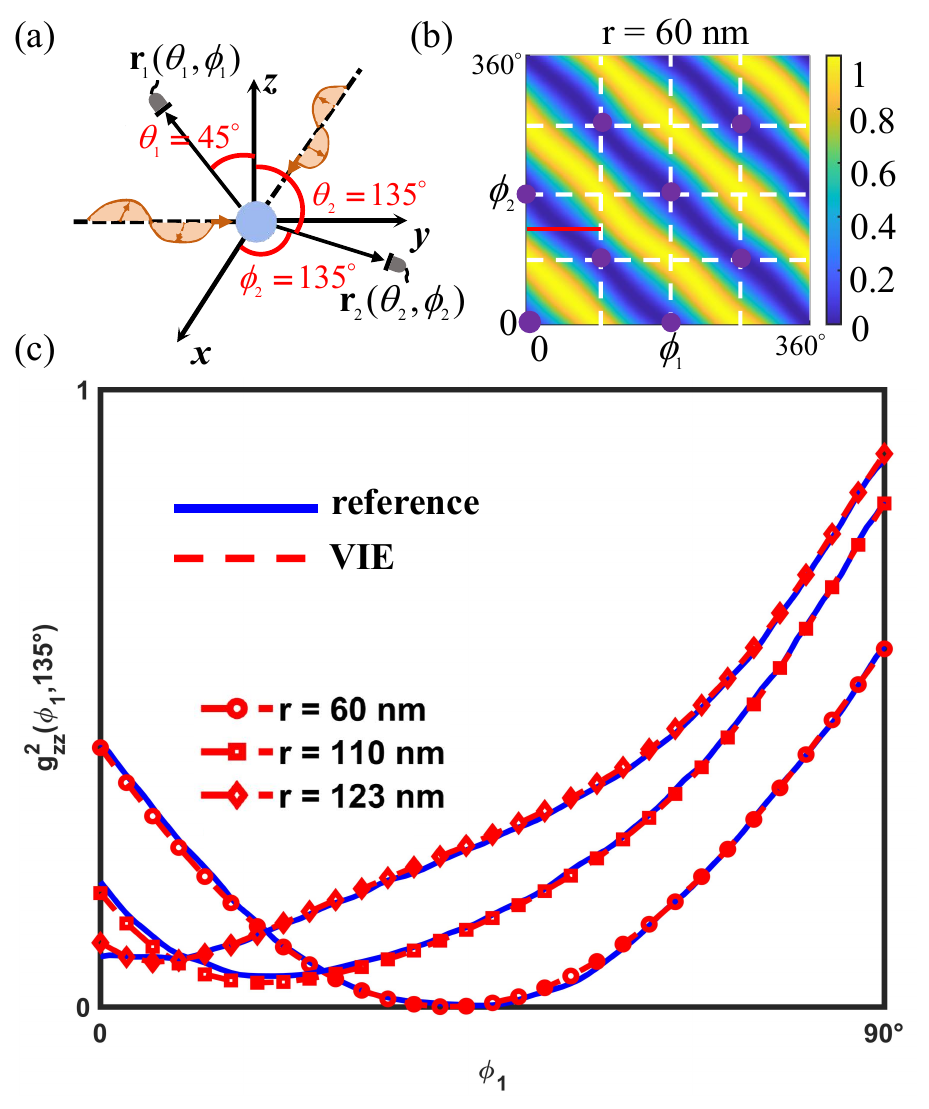}
	\caption{(a) The schematic illustrating the two-photon state and the positions of detectors with respect to the dielectric sphere. (b) Normalized second-order correlation function $g_{z z}^{(2)}\left(\phi_1, \phi_2\right)$ for the sphere at $r=60$ nm as a function of the azimuthal positions of the two detectors fixed at $\theta_1=45^{\circ},\theta_2=135^{\circ}$. (c) Normalized second-order correlation function $g_{z z}^{(2)}\left(\phi_1, \phi_2=135^{\circ}\right)$ for spheres with different radii $r$ = 60 nm, 110 nm, 123 nm.}
	\label{fig2}
\end{figure}

We validate the numerical method by comparing with the analytical results for an electrically small dielectric sphere reported in \cite{maurer2023quantum}. The sphere has relative permittivity $\varepsilon_r=2.1$ at $750~\mathrm{THz}$. The two polar angles are fixed at $\theta_1=45^{\circ}$ and $\theta_2=135^{\circ}$, and the geometry and detector configuration are shown in Fig.~\ref{fig2}(a). Using the VIE solution, the $z$-component of the far-field is computed via \eqref{eq25}, and then $g_{zz}^{(2)}$ is evaluated by \eqref{eq26}.

The red dashed curves in Fig.~\ref{fig2}(c) show $g_{zz}^{(2)}(\phi_1,\phi_2)$ for $\phi_2=135^{\circ}$ as a function of $\phi_1\in[0^{\circ},90^{\circ}]$, corresponding to the detection arc highlighted in Fig.~\ref{fig2}(b). For radii $r=60~\mathrm{nm}$, $110~\mathrm{nm}$, and $123~\mathrm{nm}$, the numerical results (red) agree well with the reference solution (blue), with relative $\ell_2$ errors of $2.81\%$, $2.73\%$, and $1.59\%$, confirming the accuracy of the implementation.

Spheres of different radii induce substantial variations in the angular dependence of $g_{zz}^{(2)}$. The inset of Fig.~\ref{fig2}(b) shows the full angular map over $(\phi_1,\phi_2)$ for $r=60~\mathrm{nm}$. The blue diagonal stripes correspond to HOM-type destructive interference, i.e., suppression of coincidences due to two-photon path indistinguishability. The wide-angle persistence of the dip is attributed to the spherical symmetry and the dominance of a small number of scattering multipoles in the electrically small regime, which makes the relative complex amplitudes in \eqref{eq26} vary smoothly with angle and hence preserves the destructive-interference condition over extended angular regions. At the two detector positions marked by the purple symbols, one of the two path amplitudes vanishes because the corresponding far-field component is identically zero for one incident channel; this yields $g_{zz}^{(2)}=0$ for a trivial single-path reason rather than genuine two-photon interference, as further clarified by the time-domain coincidence curves below.

\subsubsection{Classical fourth-order field correlation for comparison}
\begin{figure}[!t]
	\centering
	\includegraphics[width=\columnwidth]{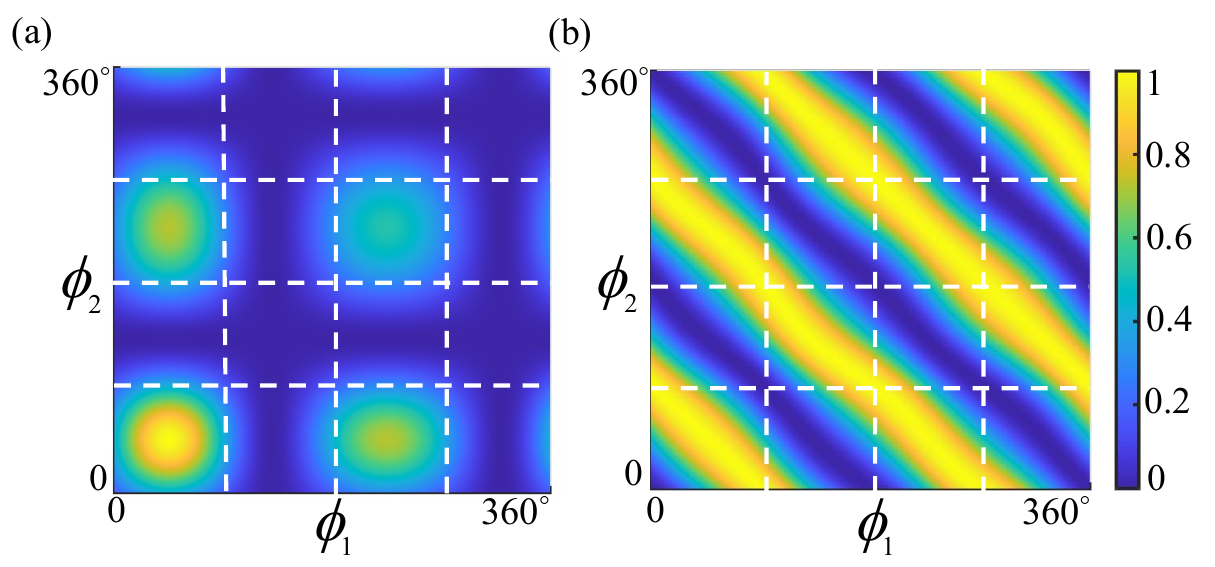}
	\caption{Angular maps of normalized fourth-order field correlations. (a) Classical intensity-product correlation $P^{(2)}_{zz}$ (normalized). (b) Quantum normalized second-order correlation $g^{(2)}_{zz}$.}
	\label{fig3}
\end{figure}

For deterministic classical fields, a natural fourth-order field correlation is simply the product of intensities,
\begin{equation}\label{eq27}
	P^{(2)}_{zz}(\mathbf r_1,\mathbf r_2)\triangleq I_z(\mathbf r_1)\,I_z(\mathbf r_2)
	=\left|E_z(\mathbf r_1)\right|^2\left|E_z(\mathbf r_2)\right|^2,
\end{equation}
where $E_z(\mathbf r)=E_{z\mathbf k_1}(\mathbf r)+E_{z\mathbf k_2}(\mathbf r)$ is the coherent superposition of the two classical scattering responses. This classical quantity can be normalized (e.g., by its maximum over the scanned angles) and compared against the quantum $g_{zz}^{(2)}$ map. Fig. \ref{fig3} contrasts the normalized classical fourth-order field correlation map with the quantum $g_{zz}^{(2)}$ map. The classical map is dominated by standard two-field interference fringes determined by second-order intensity coherence. In contrast, the quantum observable $g_{zz}^{(2)}$ encodes two-photon interference between direct and exchange amplitudes, leading to HOM-type suppression regions that have no direct classical analogue under identical normalization.

\subsubsection{Time-domain coincidence curves}
\begin{figure}[!t]
	\centering
	\includegraphics[width=\columnwidth]{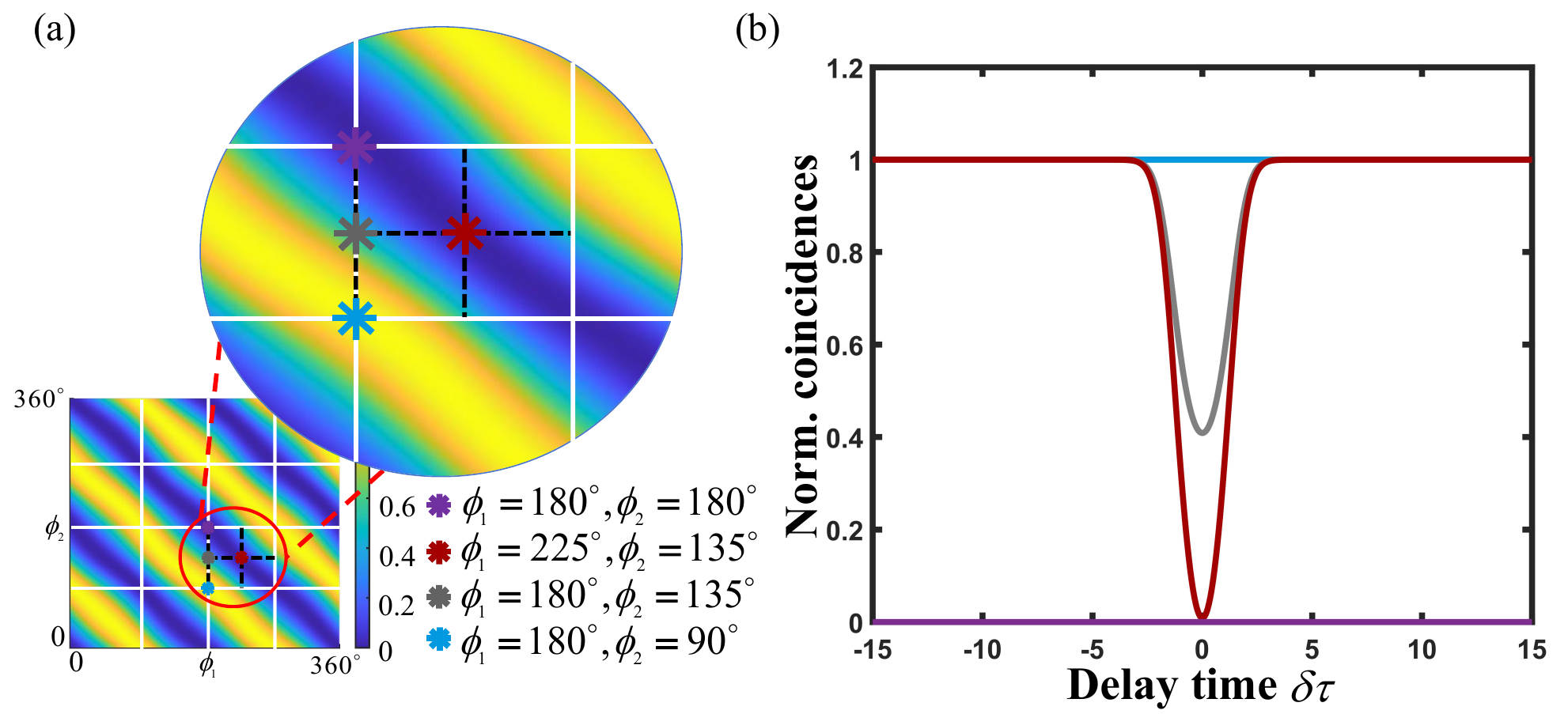}
	\caption{(a) Four representative detection configurations on the angular map. (b) Normalized coincidence counts $\tilde{N}_c$ versus the controlled input delay $\delta \tau$ for these configurations. The Gaussian width parameter used for the temporal profile is $\sigma = 1.2$.}
	\label{fig4}
\end{figure}

We further compute the normalized coincidence counts $\tilde{N}_c(\delta\tau)$ as a function of the controlled input delay $\delta\tau$ using \eqref{eq18}-\eqref{eq22}. Based on the angular map in Fig.~\ref{fig2}(b), four representative angular configurations are selected (Fig.~\ref{fig4}(a)) with substantially different values of $g_{zz}^{(2)}$. Substituting the far-field complex amplitudes at these angles into the time-domain formulation yields the curves in Fig.~\ref{fig4}(b).

The overall dip envelope is determined by the input wavepacket bandwidth (coherence time), while the dip visibility is governed by the spatially dependent two-path amplitudes in \eqref{eq26} (or equivalently, by $A_{\mathrm{dir}}$ and $A_{\mathrm{ex}}$). In particular, configurations with strong destructive interference in the frequency-domain observable exhibit deep HOM dips at $\delta\tau=0$ (red curve). When the two-photon interference is weak or absent (e.g., $g^{(2)}\approx 1$), the coincidence curve remains close to unity for all delays (blue curve). Importantly, the purple curve corresponds to a configuration where $g_{zz}^{(2)}=0$ due to the vanishing of one path amplitude (a single-photon contribution at the detector) rather than two-photon interference; accordingly, $\tilde N_c(\delta\tau)$ stays suppressed for all $\delta\tau$, distinguishing it from a genuine HOM dip whose suppression is localized near $\delta\tau=0$.

\subsection{Dielectric Metasurface}
Metasurfaces enable subwavelength control of amplitude, phase, polarization, and dispersion, with applications in flat optics \cite{khorasaninejad2016metalensesflatoptics,wang2018broadbandflatoptics}, holography \cite{huang2013threeholography}, beam shaping, and polarization control. Recently, such capabilities have also been leveraged for manipulating quantum states of light \cite{li2020metalens,stav2018quantum,yousef2025metasurface}.

Here we consider a polarization-converting metasurface based on a space-variant Pancharatnam--Berry phase (PBP) profile, designed to deflect an incident plane wave into output directions at $\pm 10^{\circ}$. For horizontally ($|H\rangle$) and vertically ($|V\rangle$) polarized inputs, the output polarization basis of interest is the left- and right-handed circular basis ($|L\rangle$, $|R\rangle$). At the operator level, it is convenient to express the corresponding mode creation operators via the circular-basis transformation:
\begin{equation}\label{eq28}
	\begin{aligned}
		\hat a_{L}^{\dagger}=\frac{1}{\sqrt{2}}\left(\hat a_{H}^{\dagger}+i\,\hat a_{V}^{\dagger}\right),
		\quad
		\hat a_{R}^{\dagger}=\frac{1}{\sqrt{2}}\left(\hat a_{H}^{\dagger}-i\,\hat a_{V}^{\dagger}\right),
	\end{aligned}
\end{equation}
where the (spatial-mode) subscripts will be specified below.

The initial two-photon state is defined in a composite Hilbert space
$\mathcal H=\mathcal H_a\otimes\mathcal H_b$, where $a$ and $b$ denote two orthogonal spatial input modes illuminating the metasurface. In this metasurface-assisted entanglement scheme, the relevant input is a polarization-entangled state in the circular basis,
\begin{equation}\label{eq:psi_in_LR}
	|\Psi_{\mathrm{in}}\rangle
	\propto\;
	\left(
	\hat a_{L,a}^{\dagger}\hat a_{L,b}^{\dagger}
	-
	e^{i\varphi}\hat a_{R,a}^{\dagger}\hat a_{R,b}^{\dagger}
	\right)|0\rangle ,
\end{equation}
i.e., a coherent superposition of the two alternatives ``both photons are left-circular'' and ``both photons are right-circular'' (the relative phase $\varphi$ depends on the controlled input delay). This state is a NOON-like entangled state in the polarization degree of freedom.

A Pancharatnam--Berry-phase (PBP) metasurface acts as a polarization-dependent router: within the dominant $\pm 1$ diffraction orders, the $|L\rangle$ and $|R\rangle$ components are deflected into two distinct far-field paths (angles) centered at $+10^\circ$ and $-10^\circ$, respectively. At the operator level, this routing can be modeled as
\begin{equation}\label{eq:router_map}
	\hat a_{L,a}^\dagger \mapsto t_L\,\hat c_{+}^\dagger,\;
	\hat a_{L,b}^\dagger \mapsto t_L\,\hat c_{+}^\dagger,\;
	\hat a_{R,a}^\dagger \mapsto t_R\,\hat c_{-}^\dagger,\;
	\hat a_{R,b}^\dagger \mapsto t_R\,\hat c_{-}^\dagger,
\end{equation}
where $\hat c_{+}^\dagger$ and $\hat c_{-}^\dagger$ create a photon in the two selected output spatial modes collected around $+10^\circ$ and $-10^\circ$, and $t_{L/R}$ are complex transfer coefficients (extracted from the full-wave simulation). Equation (\ref{eq:router_map}) assumes that the two input modes become indistinguishable within each selected diffraction channel, so that both $a$ and $b$ contribute coherently to the same output mode.

Applying (\ref{eq:router_map}) to (\ref{eq:psi_in_LR}) yields a path-NOON state in the selected far-field channels:
\begin{equation}\label{eq:psi_out_pathNOON}
		\begin{aligned}
	|\Psi_{\mathrm{out}}\rangle
	\propto\;t_L^2\left(\hat c_{+}^{\dagger}\right)^2|0\rangle
	&-e^{i\varphi}t_R^2\left(\hat c_{-}^{\dagger}\right)^2|0\rangle
	\\
	&\propto\;\left(|2_{+},0_{-}\rangle-e^{i\varphi'}|0_{+},2_{-}\rangle\right),
	\end{aligned}
\end{equation}
which is exactly the transformation ``polarization NOON $\rightarrow$ path NOON'' enabled by the metasurface-induced separation of circular polarizations. In the subsequent correlation analysis, the two detectors are aligned with the $\pm 10^\circ$ output channels, and the two-photon interference arises from the indistinguishability of the two-photon alternatives in (\ref{eq:psi_out_pathNOON}).

For evaluating the normalized second-order correlation, we compute far-field projections onto circular polarization. Using the spherical transverse components $(E_\theta,E_\phi)$, the left/right circular components are
\begin{equation}\label{eq30}
	\begin{aligned}
		E_{L}=\frac{1}{\sqrt{2}}\left(E_\theta+i E_\phi\right),
		\quad
		E_{R}=\frac{1}{\sqrt{2}}\left(E_\theta-i E_\phi\right).
	\end{aligned}
\end{equation}
Substituting the circularly polarized far-field amplitudes into \eqref{eq10} yields
\begin{equation}\label{eq31}
	\begin{aligned}
		&g_{\scriptscriptstyle {LR}}^{(2)}\left(\mathbf{r}_1, \mathbf{r}_2\right)\\
		&=\frac{\omega_{\textit{k}_H} \omega_{\textit{k}_V}\left|E_{{\scriptscriptstyle {R}} \textit{k}_H}\left(\mathbf{r}_1\right) E_{{\scriptscriptstyle {L}} \textit{k}_V}\left(\mathbf{r}_2\right)+E_{{\scriptscriptstyle {R}} \textit{k}_V}\left(\mathbf{r}_1\right) E_{{\scriptscriptstyle {L}} \textit{k}_H}\left(\mathbf{r}_2\right)\right|^2}{\sum_{l=H,V} \omega_{\textit{k}_l}\left|E_{{\scriptscriptstyle {R}} \textit{k}_l}\left(\mathbf{r}_1\right)\right|^2 \sum_{l=H,V} \omega_{\textit{k}_l}\left|E_{{\scriptscriptstyle {L}} \textit{k}_l}\left(\mathbf{r}_2\right)\right|^2}. 
	\end{aligned}
\end{equation}

\begin{figure}[!t]
	\centering
	\includegraphics[height=0.24\textwidth,width=0.8\columnwidth]{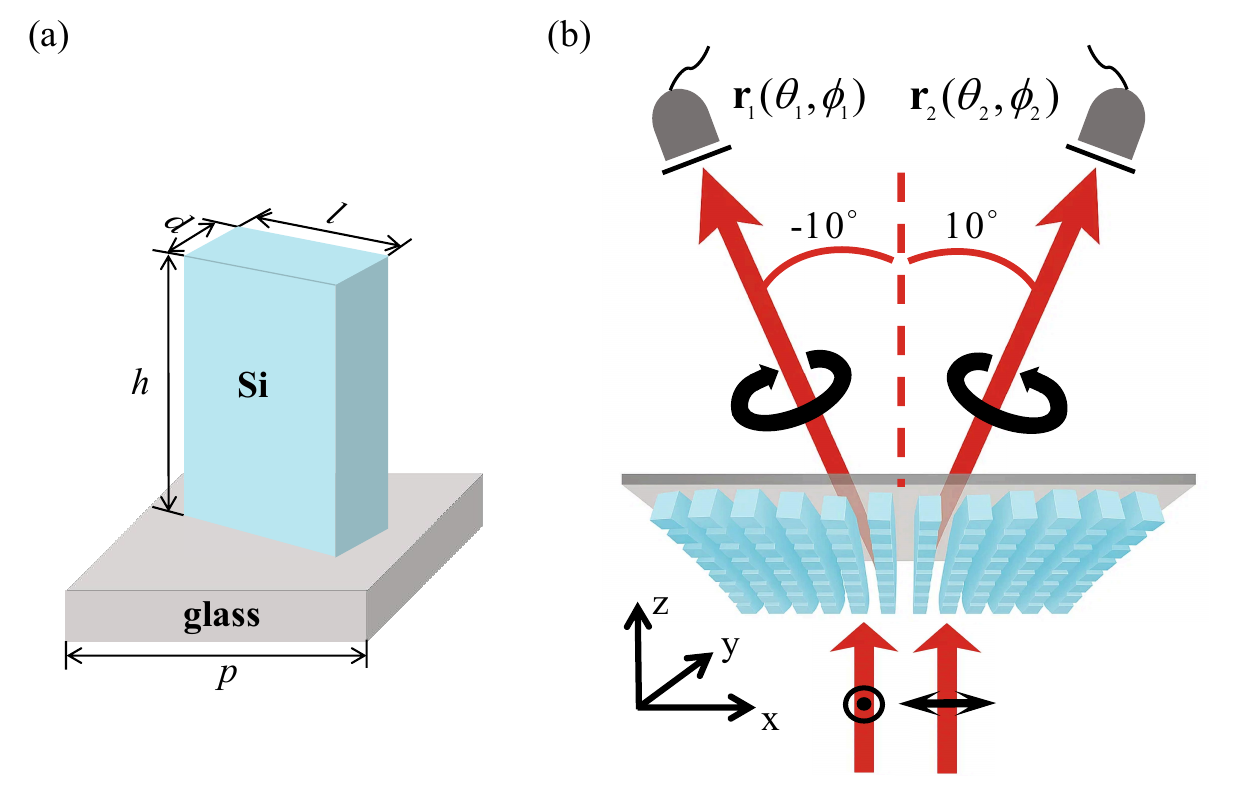}
	\caption{(a) The geometric schematic of the metasurface unit cell. (b) The schematic of the two-photon state and the positions of detectors with respect to the PBP metasurface.}
	\label{fig5}
\end{figure}

Referring to the structural parameters in \cite{georgi2019metasurface}, the PBP metasurface unit cell in Fig.~\ref{fig5}(a) uses periodicity $p=667~\mathrm{nm}$, nanofin height $h=830~\mathrm{nm}$, length $l=486~\mathrm{nm}$, and width $d=219~\mathrm{nm}$. We assume refractive indices $n_{\mathrm{Si}}=3.44$ for silicon and $n_{\mathrm{glass}}=2.25$ for the substrate at wavelength $1550~\mathrm{nm}$. The local rotation angle follows the PBP design rule:
\begin{equation}\label{eq32}
	\theta_m=\frac{1}{2} k_0 \sin (\theta_t)\, m\, p,
\end{equation}
where $\theta_t=10^{\circ}$ is the desired deflection angle. This yields a rotation increment of $13.44^{\circ}$ between adjacent unit cells. The metasurface consists of groups of 13 unit cells laterally, arranged into $7\times 7$ groups along the $x$- and $y$-directions. It functions as a circular-polarization beam separator, deflecting $|R\rangle$ toward $-10^{\circ}$ and $|L\rangle$ toward $+10^{\circ}$; the detection setup is shown in Fig.~\ref{fig5}(b).

We simulate a structure of electrical size $40\lambda_0\times 12\lambda_0\times 2\lambda_0$ with a spatial discretization of 62 grids per $\lambda_0$, resulting in $18{,}013{,}710$ mesh elements and $54{,}041{,}130$ current unknowns. The dense VIE system is solved using FFT-accelerated MVM and BiCGStab with a parallel block preconditioner. Exploiting the structural regularity, the volume is partitioned into $7\times 7$ blocks, and the preconditioned solves and FFT operations are executed in parallel on 24 threads \cite{huang2023parallel}. The preconditioner reduces the solution time from 31.9 hours to 24 hours.

The characteristic aperture dimension $D=40\lambda_0$ gives a Fraunhofer distance $R_F\approx 2D^2/\lambda_0=3200\lambda_0$, which corresponds to $\approx 4.96~\mathrm{mm}$ at $1550~\mathrm{nm}$. Since the correlation functions are evaluated at observation distances well beyond this threshold, the far-field approximation used in \eqref{eq25} is well satisfied, providing a reliable basis for evaluating $g^{(2)}$ observables.

\begin{figure}[!t]
	\centering
	\includegraphics[width=\columnwidth]{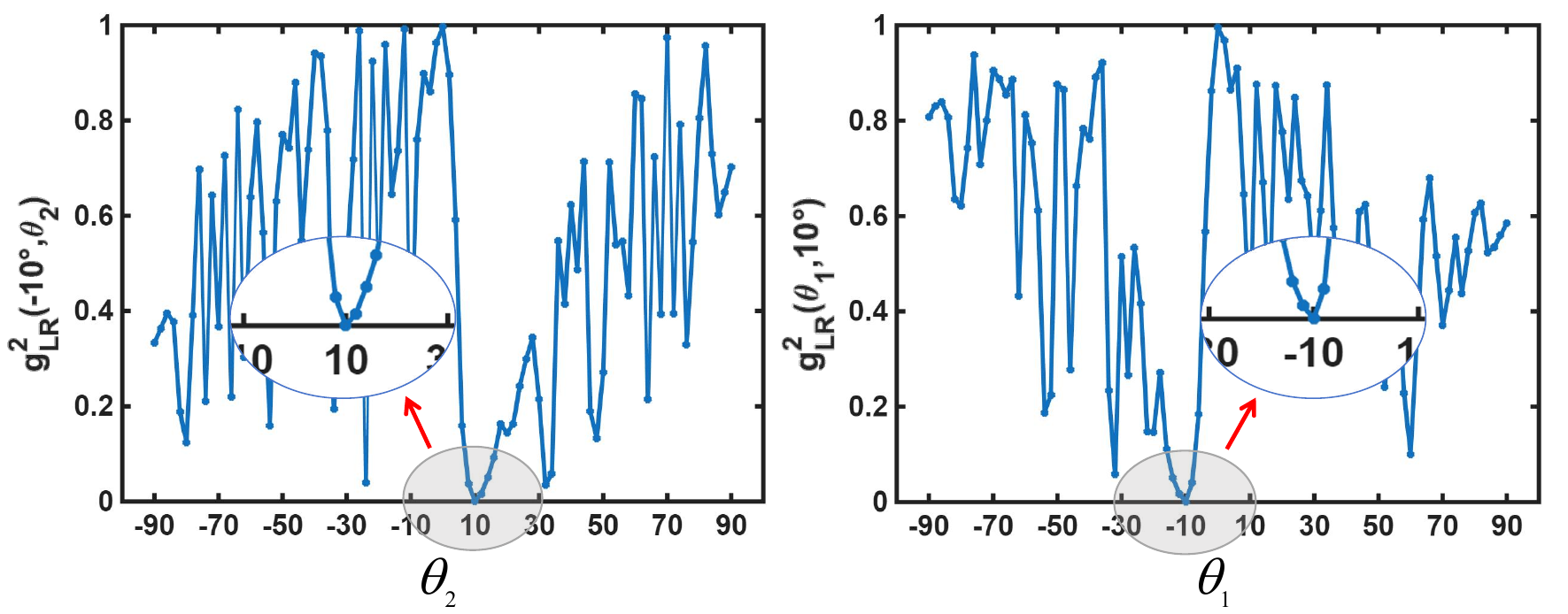}
	\caption{(a) Normalized second-order correlation function $g_{LR}^{(2)}(\theta_1=-10^{\circ},\theta_2)$ as a function of $\theta_2$. (b) Normalized second-order correlation function $g_{LR}^{(2)}(\theta_1,\theta_2=10^{\circ})$ as a function of $\theta_1$.}
	\label{fig6}
\end{figure}

Fixing the azimuthal angles at $\phi_1=180^{\circ}$ and $\phi_2=0^{\circ}$, Figs.~\ref{fig6}(a) and (b) show $g_{LR}^{(2)}(\theta_1=-10^{\circ},\theta_2)$ versus $\theta_2$ and $g_{LR}^{(2)}(\theta_1,\theta_2=10^{\circ})$ versus $\theta_1$, respectively. A vanishing correlation $g_{LR}^{(2)}=0$ occurs at $\theta_1=-10^{\circ}$ and $\theta_2=10^{\circ}$, indicating complete destructive two-photon interference (HOM suppression). The pronounced angular sensitivity of $g_{LR}^{(2)}$ highlights its utility as a two-photon probe of the metasurface scattering response.

\begin{figure}[!t]
	\centering
	\includegraphics[height=0.28\textwidth,width=0.75\columnwidth]{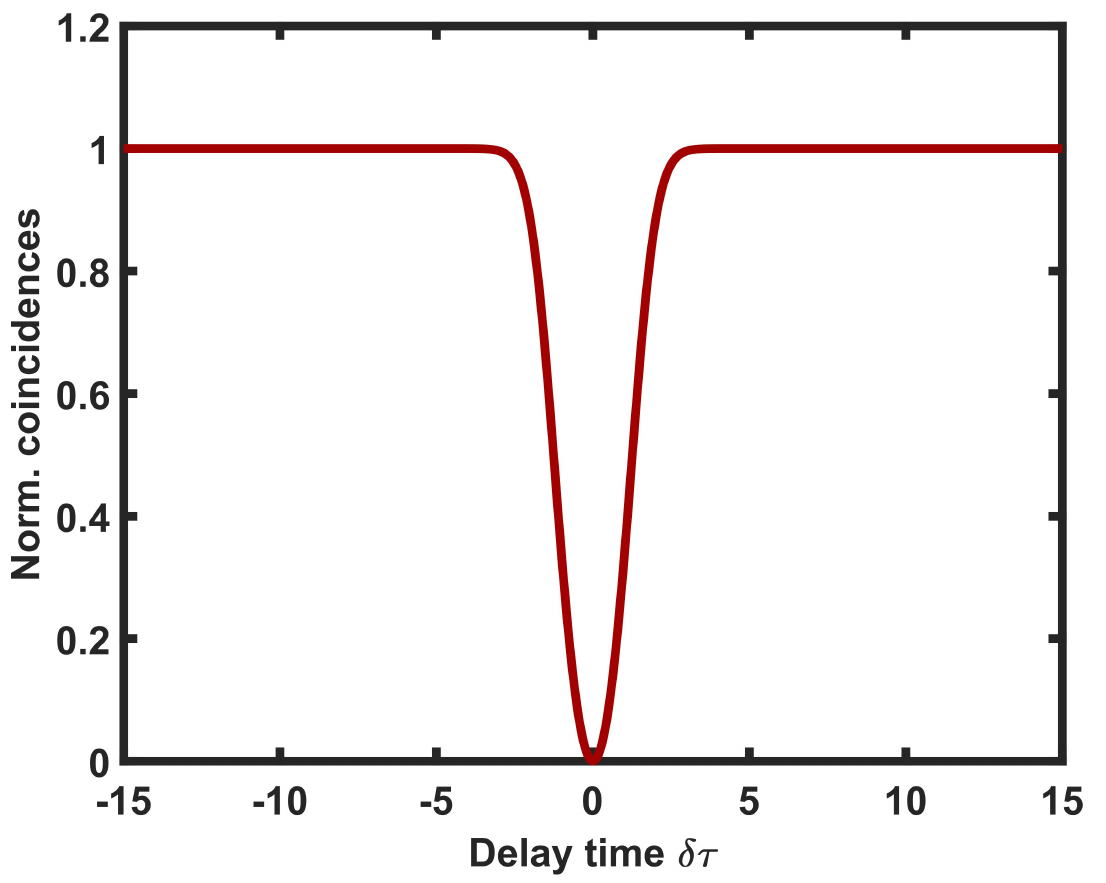}
	\caption{Normalized coincidence counts $\tilde{N}_c$ between the two detection positions versus the controlled input delay $\delta \tau$. The Gaussian width parameter used for the temporal profile is $\sigma = 1.2$.}
	\label{fig7}
\end{figure}

We also compute the normalized coincidence counts $\tilde{N}_c(\delta\tau)$ at the vanishing-correlation configuration using \eqref{eq18}. The results in Fig.~\ref{fig7} exhibit a characteristic HOM dip, consistent with the essential features reported in \cite{georgi2019metasurface}. The simulated curve is smoother than experimental traces because we assume an idealized geometry and a quasi-monochromatic Gaussian wavepacket, whereas experimental data are affected by fabrication tolerances, surface roughness, finite detector response, and background noise. Despite differences in fine-scale ripples, the dip visibility and delay dependence confirm that the proposed VIE-based framework captures the primary quantum interference behavior. As $\delta\tau$ increases and the photons become distinguishable, $\tilde{N}_c(\delta\tau)$ approaches unity, corresponding to the classical (distinguishable) limit.

\section{Conclusion}
This work investigates far-field two-photon interference associated with the HOM effect in arbitrary scattering environments. By combining a quantum photodetection formalism with full-wave computational electromagnetics, we show that the normalized second-order correlation $g^{(2)}$ and the time-domain normalized coincidence $\tilde N_c(\delta\tau)$ can be evaluated directly from classical far-field complex amplitudes computed by a VIE-FFT solver. The key observation is that, for linear media, the mapping from incident channels (direction and polarization) to far-field detection modes is linear at the operator level; therefore, once the transfer coefficients are extracted from classical scattering simulations, the two-photon observables follow from normally ordered correlations.

For a lossless scatterer, the full input-output mapping between complete sets of asymptotic modes is described by a unitary multi-channel scattering operator, which preserves the bosonic commutation relations. In our setting only two incident channels are populated and the observables are evaluated for two selected far-field detection modes; therefore the operator mapping can be projected onto this four-mode subspace, yielding an effective $2\times2$ input-to-output transfer relation whose elements are directly obtained from the full-wave computed far-field amplitudes and which leads to closed-form expressions for $g^{(2)}$ and $\tilde N_c(\delta\tau)$. The ultra-narrowband assumption is used only to treat these transfer coefficients as approximately frequency-independent over the photon bandwidth, so that the temporal-delay dependence enters solely through the wavepacket overlap function.

Numerical results for a dielectric sphere and a Pancharatnam--Berry phase metasurface demonstrate pronounced angle-dependent HOM-type suppression of coincidences. The angular distribution of $g^{(2)}$ differs fundamentally from classical fourth-order field correlations, highlighting the genuinely two-photon nature of the interference. Moreover, the frequency-domain observable $g^{(2)}$ is quantitatively linked to the visibility and shape of the time-domain coincidence dip $\tilde N_c(\delta\tau)$, allowing one to distinguish true two-photon destructive interference from trivial zero-coincidence cases caused by vanishing single-path amplitudes. We further show that $g^{(2)}$ is highly sensitive to both structural dimensions and detection angles (e.g., sphere radius variations and metasurface angular scans), suggesting that far-field correlation measurements can serve as a diagnostic modality for retrieving material/structural information and detecting subtle perturbations in electromagnetic designs.

Future work will extend this framework to lossy and dispersive systems (where the scattering description must incorporate additional bath channels), and will explore correlation-enabled quantum inverse problems for complex electromagnetic structures, including multi-angle coincidence measurements and optimization-driven design/inference using $g^{(2)}$-based objective functions.

\begin{IEEEbiography}[{\includegraphics[width=1in,height=1.25in,clip,keepaspectratio]{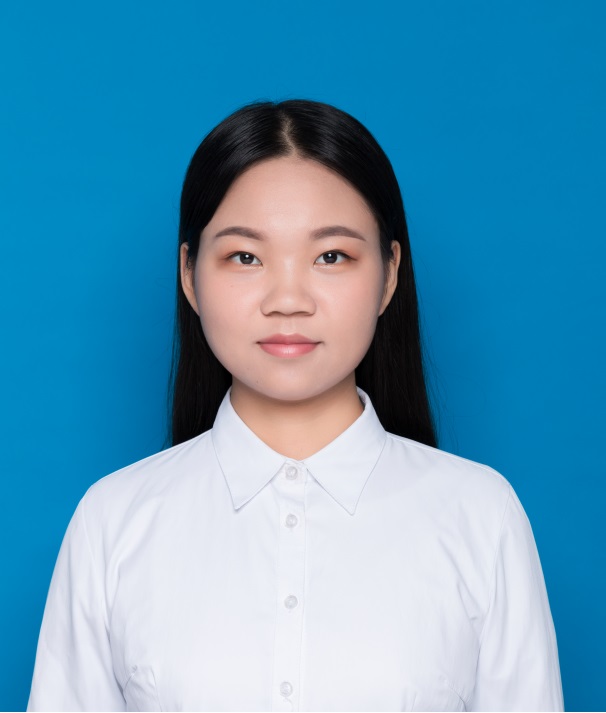}}]{Chengnian Huang} received her B. Eng. degree in Electronic Science and Technology from
	Xiamen University, Xiamen, China. in 2020. She is currently pursuing the Ph.D. degree in
	Electronic Science and Technology from Zhejiang
	University, Hangzhou, China. From Apr.
	2024 to Oct. 2024, She was a visiting PhD student
	at Department of Electrical and Electronic Engineering, the Hong Kong Polytechnic University, Hong Kong. Her research interests include the method of moments, volume integral equation method, finite-difference frequency-domain method, electromagnetic simulation of micro–nano structures, quantum scattering.
\end{IEEEbiography}

\begin{IEEEbiography}[{\includegraphics[width=1in,height=1.25in,clip,keepaspectratio]{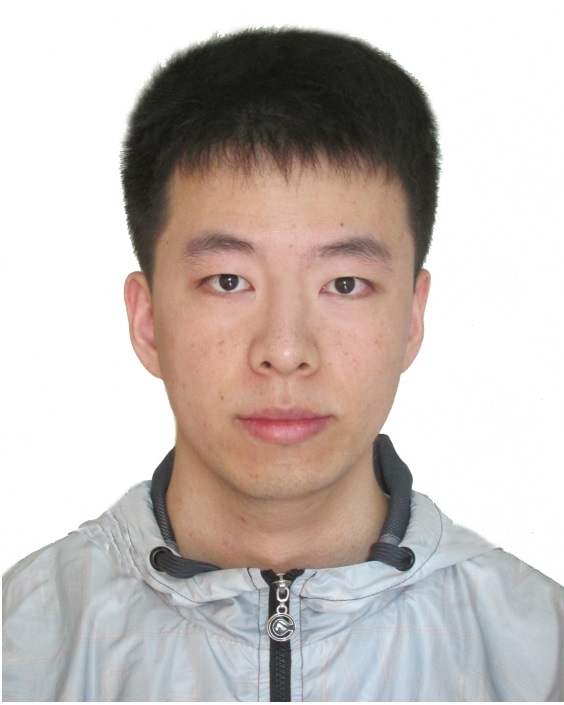}}]{Hangyu Ge} received his B.E. degree from the School of Information Science and Engineering at Shandong University in 2022. Currently, he is a PhD candidate at the College of Information Science and Electronic Engineering, Zhejiang University. His research interests include single-photon sources and entangled photon sources based on semiconductor quantum dots.
\end{IEEEbiography}

\begin{IEEEbiography}[{\includegraphics[width=1in,height=1.25in,clip,keepaspectratio]{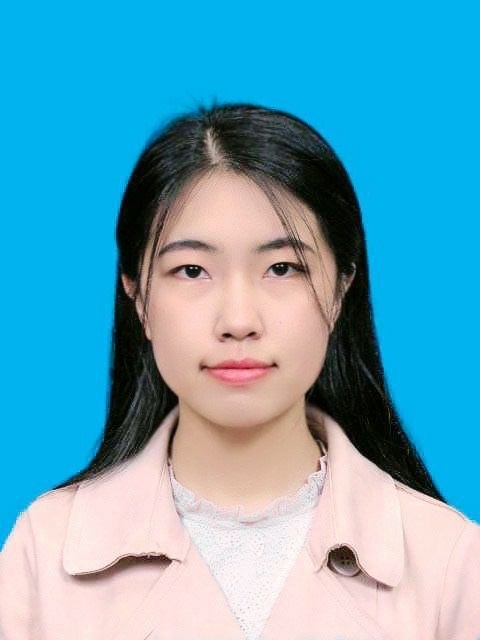}}]{Yijia Cheng} received her B. Eng. degree in Electronic Information Science and Technology from
	Chongqing University, Chongqing, China. in 2021. She is currently pursuing the Ph.D. degree in
	Electronic Science and Technology from Zhejiang
	University, Hangzhou, China. Her research interests include computational nanophotonics, scattering matrix method, fourier modal method, higher-order topological photonics.
\end{IEEEbiography}

\begin{IEEEbiography}[{\includegraphics[width=1in,height=1.25in,clip,keepaspectratio]{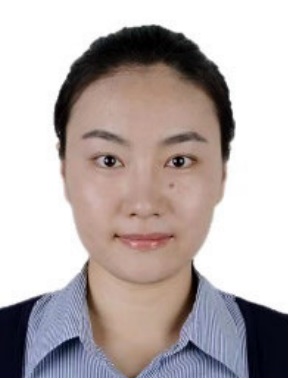}}]{Zi He} (Senior Member, IEEE)was born in
	Hebei, China. She received the B.Sc. and Ph.D.
	degrees in electronic information engineering from
	the School of Electrical Engineering and Optical
	Technique, Nanjing University of Science and Technology (NJUST), Nanjing, China, in 2011 and 2016, respectively. 
	
	She has worked as a Visiting Scholar with
	the University of Illinois at Urbana–Champaign
	(UIUC), Champaign, IL, USA, from September
	2015 to September 2016. She was a Post-Doctoral
	Researcher at the Science and Technology on Electromagnetic Scattering
	Laboratory, Beijing Institute of Environmental Features (BIEF), Beijing,
	China, from 2017 to 2020. She is currently an Associate Professor with the
	Department of Communication Engineering, NJUST. Her research interests
	include antenna, RF-integrated circuits, and computational electromagnetics.
\end{IEEEbiography}

\begin{IEEEbiography}[{\includegraphics[width=1in,height=1.25in,clip,keepaspectratio]{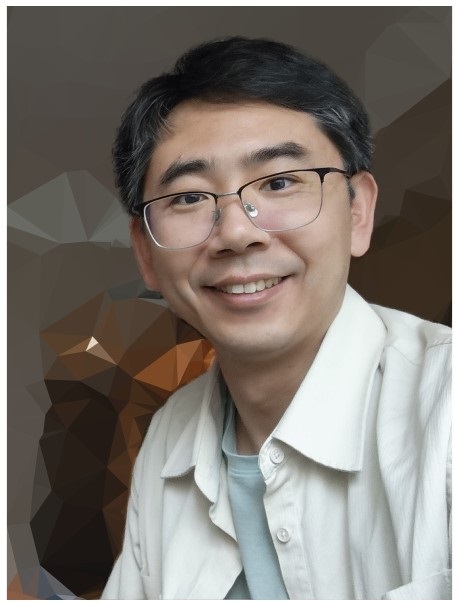}}]{Feng Liu} received his bachelor’s degree from Yantai
	University in 2008, followed by a master’s degree
	from the University of Sheffield in 2009 and a PhD
	from TU Dortmund in 2013.
	
	He conducted postdoctoral research at the Uni
	versity of Sheffield and RWTH Aachen from 2013
	to 2019. In 2019, he joined the College of In
	formation Science and Electronic Engineering at
	Zhejiang University, where he is currently a tenured
	associate professor. His research focuses on quan
	tum light sources based on III-V Quantum Dots,
	cavity/waveguide-QED, and integrated quantum optical circuits.
\end{IEEEbiography}

\begin{IEEEbiography}[{\includegraphics[width=1in,height=1.25in,clip,keepaspectratio]{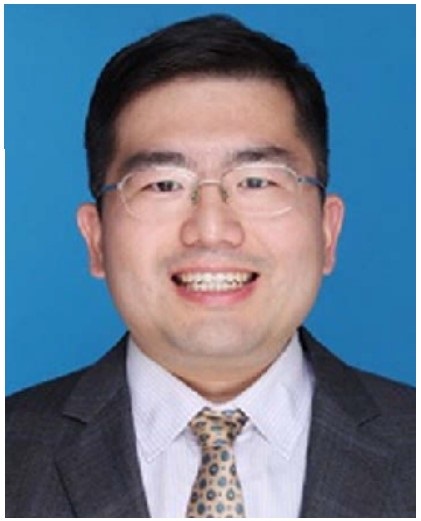}}]{Wei E. I. Sha} (Fellow, IEEE)  received his B.S. and Ph.D. degrees in Electronic Engineering from Anhui University, Hefei, China, in 2003 and 2008, respectively. From July 2008 to July 2017, he served as a Postdoctoral Research Fellow and later a Research Assistant Professor in the Department of Electrical and Electronic Engineering at the University of Hong Kong. Between March 2018 and March 2019, he was a Marie Skłodowska-Curie Fellow at University College London (UCL), UK. In October 2017, he joined the College of Information Science and Electronic Engineering at Zhejiang University, Hangzhou, China, where he is currently a tenured Associate Professor and doctoral supervisor, and Deputy Director of the Institute of Electromagnetic Information and Electronic Integration Innovation.
	
	Dr. Sha’s long-term research focuses on the fundamental theories and engineering applications of computational electromagnetics, nano/quantum electromagnetics, and electromagnetic information. He has authored or co-authored more than 220 peer-reviewed SCI-indexed journal articles. He has delivered 7 keynote speeches, 3 short courses, and 40 invited symposium presentations at international conferences. His research work has been cited over 13,000 times on Google Scholar, with an h-index of 60.
	
	Dr. Sha is a Fellow of the Institute of Electrical and Electronics Engineers (IEEE). He has reviewed for numerous top technical journals in the field and served on the Technical Program Committees of many international academic conferences. He also holds editorial positions in several prestigious academic journals related to electromagnetics and antennas.
	
	Dr. Sha has received a series of academic awards, including the Technical Achievement Award from the Applied Computational Electromagnetics Society (ACES), the Second Prize of the Science and Technology Award from the China Institute of Communications, and the Second Prize of Anhui Provincial Science and Technology Award. Nine of his students have won Best Student Paper Awards at international and domestic academic conferences.
\end{IEEEbiography}

\vfill

\end{document}